\documentstyle[eqsecnum,aps]{revtex}

\scrollmode

\tightenlines

\begin{document}

\def\Kbar{\overline{\!K}}

\draft

\title{Three-Body approach to the $K^-d$ Scattering Length in Particle Basis}

\vspace{0.5cm}

\author{A. Bahaoui$^1$, C. Fayard$^2$, T. Mizutani$^3$, and B. Saghai$^4$}

\address{1) 
Universit\'e Chouaib Doukkali, Facult\'e des Sciences,
El Jadida, Morocco} 

\address{2) Institut de Physique Nucl\' eaire de Lyon, IN2P3-CNRS,
 Universit\'e
Claude Bernard,\\ F-69622 Villeurbanne Cedex, France} 

\address{3) Department of Physics,
Virginia Polytechnic Institute and State University,\\
Blacksburg, VA 24061 USA}

\address{4) Service de Physique Nucl\' eaire, DSM/DAPNIA, CEA/Saclay, \\
F-91191 Gif-sur-Yvette, France}

\date{\today}

\maketitle

\begin{abstract}
We report on the first calculation of the scattering length $A_{K^-d}$  
based on  a relativistic three-body 
approach where the two-body input amplitudes coupled to the $\Kbar N$ 
channels have been obtained with the chiral $SU(3)$ constraint, 
 but with
isospin symmetry breaking effects taken into account.  
Results are compared  
 with a recent calculation applying a similar set of two-body amplitudes,  
based on the fixed center approximation, considered
as a good  approximation for 
a loosely bound target, and for which we find significant deviations 
from the exact three-body results. Effects of the hyperon-nucleon 
interaction,  
and deuteron $D$-wave component are also evaluated.
\end{abstract}



While the threshold behavior of the $K$-nucleon system has been found simple, 
the corresponding one for the
$\Kbar$-nucleon ($\Kbar N$) is quite complicated as its threshold is above 
those for the 
$\pi Y$ ($Y \equiv \Lambda, \Sigma$) channels to which it couples 
strongly \cite{eisen}.
Besides, it also couples to the below-threshold $\Lambda (1405)$ resonance.  
Moreover, this topic has suffered from years of persistent ambiguity in the 
sign of the real part 
of the $K^-p$ scattering length $a_{K^-p}$: sign from the scattering data 
opposite to the one 
from kaonic hydrogen atomic data. 
Even under this circumstances, a few three-body calculations on the 
$K^-$- deuteron 
scattering length
$A_{K^-d}$ were performed with different degrees of refinement, 
by always disregarding the controversial kaonic hydrogen constraint on  
$Re(a_{K^-p})$ \cite{hetherington,schick,toker,torres,baha_paper,baha_thesis}.
Some of these works were devoted primarily to the calculation of the  mass 
and momentum  distributions such as  $m(\pi Y)$, in the breakup reactions: 
$K^-d \to \pi NY$, so the $K^-d$ scattering length was, to some extent, 
a by-product  
\cite{toker,torres}. Calculations required various two-body  amplitudes 
as input, 
the most important of 
which being the coupled $\Kbar N, \ \pi Y$ channels.  Those amplitudes 
were derived
from  {\it ad-hoc} rank one separable potentials with  (energy independent) 
strengths, 
and ranges in the form factors determined  by fit to the low energy $K^-p$ 
scattering 
data.  On the 
average the thus obtained values for $A_{K^-d}$  were  centered around 
$\approx (-1.5 + i1.0)\ \hbox{fm}$.  Due to the very restricted quantity and 
quality of 
the data and to the lack of sound theoretical guidance 
(apart from isospin symmetry) on the form 
of the potentials, along 
with the then troubled $Re(a_{K^-p})$, it appeared meaningless to continue this 
theoretical  
endeavour any further. So, the investigation in the subject became dormant. 
One very important finding, however, was that the iterative solution for $K^-d$ 
did diverge, hence solving 
the three-body equations without truncation became a must.  

Recently, there has been a steady progress in effective low energy hadronic 
methods such as 
Chiral Perturbation Theory  \cite{meissner,ecker}.  
This advance as well as the new $K^-p$ data have created a renewed interest  
in the physics with low energy kaons, to the extent that 
there have even been discussions on extracting  the kaon-nucleon $\sigma$ terms 
which 
are expected to provide important information on chiral symmetry breaking,  
strangeness 
content of the nucleon, etc. \cite{guaraldo,olin,gensini}. Note that both  
$a_{K^-p}$ and $A_{K^-d}$ are vital ingredients in this respect.

On the experimental side, the long-standing  
sign puzzle in $a_{K^-p}$ got finally resolved  by the KEK  X-ray measurement 
in the 
kaonic hydrogen  \cite{Iwa}.  The extracted scattering length is:
$a_{K^-p} =      (-0.78 \pm 0.15  \pm 0.03) 
           + i   (-0.49 \pm 0.25  \pm 0.12)\ \hbox{fm}$.	        
Though the sign of the real part is now settled, one clearly needs a more 
accurate value, particularly for its imaginary part.  With this in mind, 
remeasuring this quantity 
along with extracting $A_{K^-d}$ from  kaonic atom experiments is 
underway in DEAR experiment at DA$\Phi$NE, see e.g. \cite{guaraldo}. 
This should, 
in principle, allow for an extraction  of the scattering length $a_{K^-n}$ 
(see e.g. Ref. \cite{barrett}).  

Interest in improving the calculation of $A_{K^-d}$ may be witnessed in two
recent publications. 
First, Deloff \cite{deloff} compared the results of old generation 
multi-channel three-body calculations \cite{torres} with a simplified 
three-body result keeping only the 
$K^-p$ and $K^-n$, and $NN$(deuteron) input (all in $S$-wave), and with  the 
fixed centre approximation (FCA) applied to the simplified three-body model. 
Here the positions 
of the proton and neutron in the target deuteron were frozen at a  
certain separation while the $\Kbar N$ amplitudes were replaced by their 
scattering lengths. The $K^-d$ amplitude was then obtained  algebraically 
as a function 
of the proton-neutron separation. To include partially the effect of 
Fermi motion,
its expectation value over the separation was calculated with the deuteron 
wave function.
(this leads to the results called {\it FCA-integ} in \cite{deloff}).
Second, Kamalov et al. \cite{kamalov} performed yet another FCA calculation, 
but with 
an essential difference: the input $\Kbar N$ potentials for the $S$-wave 
amplitudes were obtained at $O(1/f^2)$,
lowest order in the  $SU(3)$ non-linear chiral Lagrangian for the 
interaction of the
pseudoscalar meson and $1/2^+$ baryon octets \cite{osetramos}. Only two 
free parameters were involved: the best fit to the data was found with a  
cut-off in 
the momentum integration at $p_{max}=660$ MeV, and with an 
effective  meson decay constant $f$ only $~15\%$ larger than the physical pion 
decay constant:
$f_{\pi}=93$ MeV. 
With the {\it hadron physical masses} resulting from the isospin symmetry 
breaking, 
which the authors called the 
{\it Physical} (or {\it Particle}) {\it Basis} as compared with the 
{\it Isospin Basis}, 
the obtained amplitudes for the coupled 
$\Kbar N, \ \pi Y,\ \eta Y$ channels allow to reproduce the existing 
low energy data  quite well (for References, see \cite{osetramos}).  
The $\Lambda(1405)$ resonance was also 
generated as a bound state below the $K^-p$ and $\Kbar ^\circ n$ thresholds. 
This approach is in sharp contrast 
to the models mentioned earlier, in which the only constraint on the 
numerous parameters was the $\chi ^2$ fit to the available cross sections, 
etc.  
(note, however, that improved models of this type exist with $SU(3)$ 
constraints on the relative strengths of the potentials \cite{siegsag}). 

Here we have chosen to employ a strategy similar to the one in  
Ref.~\cite{osetramos} 
for determining the essential part of the 
input to the three-body equations, and solve them exactly. In this way, 
we will be able not only to provide the best theoretical value 
for $A_{K^- d}$ to date, but also to 
test the reliability of the FCA, the effect of the $\pi N$ and $YN$ 
interactions, etc. on 
this quantity, as investigated in \cite{deloff} within the old scheme.

We have introduced  two distinct sets of potentials which are slightly 
different from the one 
in \cite{osetramos}.  The main reasons are: (i) to  check the sensitivity 
of the calculated $A_{K^- d}$  to the two-body input  within a reasonable 
margin of difference, 
and (ii) to embody them in our current investigations on the finite energy 
$K^- d$ scattering 
including the three particle final states like $\pi NY$, for which the 
momentum integration 
must be done along a rotated line in the complex plane.  For this objective, 
instead of truncating
the integration at $p_{max}$, the potentials should have a 
smooth cut-off by 
form factors. Following closely Eqs. (1) to (9) of  \cite{osetramos}, the first 
set of potentials ($\equiv$  OS1) is expressed, using the isospin notation, as:
$$V(I)_{ij}\equiv -\frac{1}{4f^2}C^I_{ij}g(p_i)(\epsilon_i +\epsilon_j)g(p_j),$$
where $p_i$ and $\epsilon_i$ are the magnitude of the center of mass momentum 
and the 
corresponding meson energy in the $i$-th channel, respectively. The $SU(3)$ 
coupling 
coefficients are $C^{I=0}_{ij}\equiv D_{ij}$  
and , $C_{ij}^{I=1}\equiv F_{ij}$, as defined in Tables II and III 
of~\cite{osetramos}. 
The form factor has 
been chosen as $g(p)=\beta^2 /(p^2+\beta^2)$ for all the coupled channels.  
A fit to the data with comparable quality to  \cite{osetramos} has been reached 
with $\beta=870$ MeV (4.41 fm$^{-1}$) and  
$f=1.20f_{\pi}$.  The second set of potentials ($\equiv$  OS2)  introduces the 
possible $SU(3)$ breaking effect in the coupling strengths such that 
its form is 
identical to the one for OS1, except that it is now multiplied by an extra 
coefficient $b^I_{ij}$.  
By performing a standard statistical fit to the data, we have obtained 
$\beta=865$ MeV (4.39 fm$^{-1}$) and $f=1.16f_{\pi}$. 
The values of the $SU(3)$ breaking coefficients all stay within $20\%$ around 
unity, see Table \ref{tab:model-OS2}.  Note that, unlike in \cite{siegsag}, 
the radiative capture $K^-p \to \gamma Y$ has not been investigated.  Overall, 
the fit to data by these two 
interactions and the one in \cite{osetramos} are just about the same: 
differences may be examplified in terms of the scattering lengths shown 
in Table \ref{tab:AKN}. 
All of them have been evaluated at the $K^-p$ threshold (=1432 MeV): beware the 
discussion below regarding the value of the
threshold at which these quantities are calculated. As compared with 
experiment \cite{Iwa},
both the real and imaginary parts of $a_{K^-p}$ given by all models adopted
are found within $2\sigma_{stat}$  
of the central values. The extra parameters in OS2 make the results somewhat 
distinct from the two other models. The symmetry  breaking effect in the mass 
of the hadron isospin multiplets on the scattering lengths is quite visible, 
especially on the real parts, as one can see in Table \ref{tab:AKN}. 
(In the limit of isospin symmetry, one has $a_p=a_n^\circ$,
and $a_{ex}=a_p-a_n$). 
Finally, we should note that just like in \cite{osetramos} we have retained
also the 
$\eta Y$ channels to obtain a reasonable fit to some data like the 
$\pi \Sigma$ mass spectrum.

Other two-body input for our three-body equations consists of the $NN$ 
interaction in the 
deuteron channel, and the $P_{33}$ $\pi N$  
and $S$-wave $YN$ interactions. The first one not only holds the initial- and 
final- state proton and neutron to form a deuteron, but supplies the $NN$ 
scattering in the 
presence of a spectator kaon in intermediate processes. We have adopted a 
specific model of 
our choice, but taken also two other models for comparison, 
as discussed later.  
The remaining interactions have been taken from \cite{baha_paper,baha_thesis}.

In our three-body calculation, we first retain two-body  $\Kbar N$ 
$t$-matrices only:  for the elastic $K^-p,\ K^-n,\ \Kbar^\circ n$, and charge 
exchange
$K^-p \leftrightarrow  \Kbar^\circ n$, which is in line with \cite{kamalov}.  
It turns out that,  
with only these two-body channels for $K^-d$ at threshold, effectively 
there is no other 
strong branch cut along the real axis 
in the momentum integration in the three-body equations, so no contour 
rotation into 
the complex plane is needed for integration, and even a sharp cut-off 
may be imposed. 
Thus with the two-body input 
from  \cite{osetramos}, we were able to find the exact solution to the 
three-body 
equations $without$ making the FCA as adopted in \cite{kamalov}.  
Table \ref{tab:AKd_FCA} summarizes our calculations. The result with  
the amplitudes 
from \cite{osetramos} is presented as in column Oset-Ramos, along with  
our own sets of  
two-body input OS1 and OS2.  For later discussions we have separated 
the results into: 
(i) pure elastic case: $K^-$ multiple scattering on the proton and neutron,
(ii) the total contribution,
(iii) the intermediate charge exchange contribution, which is the 
difference between
the values in (ii) and (i).   

We first present the consequences resulting from the {\it on-shell} properties 
of the $\Kbar N$ input. Although not entirely free from ambiguity, one possible
way to 
define  the on-shell contribution may be given by the corresponding FCA result.
Given that the deuteron is very loosely bound, this could in fact 
be guaranteed by B\' eg's theorem \cite{beg}: if the ranges of  
interactions for the 
projectile and target constituents between two successive  collisions do 
not overlap, 
the projectile-target interaction is described entirely by the on-shell 
properties of the 
two-body input. From Table \ref{tab:AKd_FCA}, we see that the results for 
all three models
are more or less the reflection of the differences in the scattering lengths 
in Table \ref{tab:AKN}. 
Now there is a bit of trouble in the present situation: near the threshold  
the $K^-p$ and  $\Kbar^\circ n$ elastic, and $K^-p \leftrightarrow \Kbar^\circ n$ 
charge exchange amplitudes all vary rapidly due to the proximity of the 
$\Lambda (1405)$ resonance. In fact, the minimum of the real part of 
the ${K^-p}$ amplitude is found 
to be located slightly below the  $K^-p$ threshold ($W_{th}=1432$ MeV), 
see e.g. Fig. 9 of \cite{osetramos}.  Besides, the threshold is slightly 
different for each 
physical $\Kbar N$ channel, except in the limit of exact isospin symmetry. So, 
depending on the threshold energy adopted in determining the scattering lengths 
$a_{K^-p}$, $a_{\Kbar^\circ n}$ and  $a_{K^-p \leftrightarrow \Kbar^\circ n}$ 
for use in the FCA, the resulting {\it on-shell} contribution to 
$A_{K^-d}$ has been 
found to vary 
up to at least $20\%$ for its real part. On the other hand, 
its imaginary part is relatively stable. It may be useful to remark that this 
strong variation in the present FCA result
is due to the violation of B\' eg's theorem: the finite life time of the 
$\Lambda (1405)$
causes its propagation, hence  non-overlapping of the interaction ranges does 
not materialize. 

Now, we wish to underline a significant finding of the present work: 
as one can see 
in Table \ref{tab:AKd_FCA}, the Faddeev results for all three models are closer 
to each 
other than in FCA,
with $Re(A_{K^-d}$) for OS2 only about 15$\%$ different from the values given
by the other two models.   
By comparing the exact three-body result and its FCA version for a given set 
of two-body 
$\Kbar N$ interaction,  there is a noticeable difference which may be
regarded as due to off-shell effects.  Particularly,  the effect of the charge 
exchange scattering $K^-p \leftrightarrow \Kbar^\circ n$ in multiple scattering 
process has 
been found grossly overestimated in the FCA. This is because the  $\Kbar^\circ n$ 
channel has a higher threshold than  that for $K^-p$. The constant 
scattering length 
approximation adopted in FCA ignores this aspect. Within the  FCA, 
the situation gets even 
worse with the $Isospin\  Basis$ in which the two thresholds are identical: 
see e.g. Table II of \cite{kamalov}.   

Next we have checked the dependence on deuteron models. First, we have compared 
the result with 
our~\footnote{The parameters are fitted to the static properties of the 
deuteron,
with $D$-state percentage value $P_D=6.7 \%$, and to the monopole charge 
form factor up to $\sim 6$ fm$^{-1}$.} SF(6.7) deuteron \cite{d_giraud} 
with the one using the relativized version of the model 
elaborated in \cite{d_pest1}. 
The difference in $A_{K^-d}$ was found mostly in the imaginary part, but was 
only within a few percent. But when a simple $^3S_1$-wave
model is used, this difference grows to be about $20\%$ as seen in the second 
and third columns of Table \ref{tab:AKmd_2}. However, the real part appears 
quite stable. The short range part of the deuteron wave function should be 
responsible for this difference, and at least one needs to retain a realistic 
model with the $^3D_1$ component.  In the same Table, we give a comparison 
between the results in the {\it Particle} and {\it Isospin} bases.  
The difference is remarkable in the imaginary part, 
which  may be easy  to understand since all the $\Kbar N$ thresholds are 
identical in the {\it Isospin Basis}, 
hence charge exchange scattering is kinematically "elastic", as discussed 
in the last paragraph.  

We then want to check the claim in \cite{deloff} that the FCA 
is rather reliable relative to the full three-body result. In fact, 
by comparing the rows for
{\it FCA-integ} and {\it Faddeev} in Table II of \cite{deloff}, 
the author seems to be right:
the two methods provide almost identical imaginary parts, while the FCA tends
to slightly underestimate the magnitude of the real part. This is just opposite 
to what has been 
found above: see Table \ref{tab:AKd_FCA}. Eventually, we have solved this 
apparent puzzle: by taking a pure $S$-wave deuteron and 
also by excluding the charge exchange contribution in the $\Kbar N$ input to 
the three-body 
equations, we have found that the exact and FCA solutions present very similar 
values for the 
imaginary part, but that the latter underestimates 
the real part by about $30\%$. In fact this is how the author of 
\cite{deloff}  performed his calculation, and the 
characteristic of the outcome was just the same: main difference in 
the real part. 
Then, once 
the charge exchange contribution is introduced, we find that the trend changes 
considerably.  We have found further that by introducing a
realistic deuteron with the $D$-component, even the result without charge 
exchange process 
does not satisfy the finding of \cite{deloff}.  Hence we conclude that 
the FCA is not as 
reliable as claimed in \cite{deloff}.

Lastly, we need to check the effects due to the $\pi N$ and $YN$ interactions, 
which have been excluded 
so far from our two-body input: they introduce the $\pi (YN)$ and $Y(\pi N)$ 
states in the three-body equations, where particles outside the parenthesis are 
the spectators. Our preliminary results show effects smaller than $5\%$, 
so the semi-quantitative estimate of \cite{kamalov} seems justified, thus the 
claim towards the end of  \cite{deloff} appears on no solid ground 
at this point.

To summarize, starting from a study of $K^- p$ scattering length, reproducing 
well enough the data, we presented a relativistic Faddeev approach for the
$K^-d$ scattering length and investigated its sensitivity to various input
ingredients. The obtained values for $A_{K^-d}$ agree with each other within
$\pm$20\%, leading to $A_{K^-d} \approx (-1.8 + i1.5)\ \hbox{fm}$. 
Here, our approach embodied elastic and inelastic $\Kbar N$ channels
in the three-body formalism.
To go further, we are in the process of including all other relevant inelastic
channels, such as $\pi Y$ and $\eta Y$. How one may extract the scattering 
length $a_{K^-n}$ from the experimental values of $a_{K^-p}$ and $A_{K^-d}$, 
is an other question under study. 
A more extensive account will be reported in a forthcoming paper.

\vskip 0.2 cm

A.B. and T.M. want to thank IPN, Lyon for kind hospitality extended to them 
in the course of this enterprise. A.B. also thanks CEA, Saclay for a
generous six months hospitality. We are indebted to A. Ramos,  R. Machleidt 
and A. Olin for their kind help in several issues related to the present work.  

%


\newpage

\begin{table}
\caption{SU-(3)-symmetry breaking coefficients $b^I_{ij}$ $(\equiv b^I_{ji})$ 
for model OS2.} 
\protect\label{tab:model-OS2}

\medskip

\begin{center}
\begin{tabular}{ccccccccc}
%
   I=0  & $\Kbar N$ & $\pi \Sigma$ & $\eta\Lambda$
	  &  I=1
          & $\Kbar N$ & $\pi \Sigma$ & $\pi\Lambda$  & $\eta\Sigma$\\
%
\hline
%
   $\Kbar N$  & $0.93$ & $1.19$ & $0.84$
 & $\Kbar N$  & $1.07$ & $1.20$ & $0.83$& $1.07$\\
%
   $\pi \Sigma$   &   & $0.87 $  & $0$ 
 & $\pi \Sigma$   &   & $0.81 $  & $0$   & $0$ \\
%
   $\eta\Lambda$  &   &       & $0$
 & $\pi \Lambda$  &   &       & $0$     & $0$   \\
%
                  &   &       & 
 & $\eta \Sigma$  &   &       &       & $0$  \\
%
%
\end{tabular}
\end{center}
\end{table}

\begin{table}
\caption{$\Kbar N$ scattering lengths (in fm) calculated at $W=M_{K^-}+M_p$ 
in the particle basis with models OS1 and OS2. The values in the last column
have been evaluated by Ramos~[17] at the same energy.
$a_p$, $a_n$, $a_n^0$, and $a_{ex}$ are the scattering lengths for 
elastic $K^-p$, $K^-n$, $\Kbar^\circ n$, and charge exchange 
$K^-p \leftrightarrow  \Kbar^\circ n$, respectively.}
\protect\label{tab:AKN}

\begin{center}
\begin{tabular}{cccc}
%
   & OS1  &    OS2   &  Oset-Ramos \\
\hline
$a_p$  & $ -1.04 + i\, 0.83 $ 
                    & $ -0.71 + i\, 0.92 $ & $ -1.01 + i\, 0.95 $ \\            
$a_n$  & $ \phantom{-} 0.57 + i\, 0.45 $ 
                    & $ \phantom{-} 0.71 + i\, 0.69 $ 
		                    & $ \phantom{-} 0.54 + i\, 0.53 $ \\
$a_n^0$ 
                      & $ -0.60 + i\, 0.89 $ 
                      & $ -0.23 + i\, 0.97 $ & $ -0.52 + i\, 1.05 $ \\
$a_{ex}$  
                      & $ -1.37 + i\, 0.48 $ 
                      & $ -1.16 + i\, 0.39 $ & $ -1.29 + i\, 0.48 $ \\
\end{tabular}
\end{center}
\end{table}

\begin{table}
\caption{$K^- d$ scattering length (in fm) calculated in the particle basis,
with the FCA approximation, and with the Faddeev three-body model.
The FCA and the Faddeev calculations in the last column have been 
performed by us 
with the Oset-Ramos $\Kbar N$ scattering lengths given in Table \ref{tab:AKN}.}
\protect\label{tab:AKd_FCA}

\begin{center}
\begin{tabular}{cccc}
 FCA   &   OS1  &    OS2   &  Oset-Ramos  \\
\hline
el. only          & $ -1.32 + i\, 1.10 $ 
                  & $ -1.09 + i\, 1.41 $ & $ -1.36 + i\, 1.26 $ \\            
charge ex.        & $ -0.83 + i\, 0.82  $ 
                  & $ -0.64 + i\, 0.35 $ & $ -0.63 + i\, 0.69 $ \\
total             & $ -2.15 + i\, 1.92 $ 
                  & $ -1.73 + i\, 1.76 $ & $ -1.99 + i\, 1.95 $ \\
\hline
 Faddeev  &     &       &    \\
\hline
el. only          & $ -1.70 + i\, 1.31 $  
                  & $ -1.41 + i\, 1.48 $ & $ -1.68 + i\, 1.33 $ \\ 
charge ex.        & $ -0.29 + i\, 0.34 $ 
                  & $ -0.27 + i\, 0.18 $ & $ -0.24 + i\, 0.25 $\\
total             & $ -1.99 + i\, 1.65 $ 
                  & $ -1.68 + i\, 1.66 $ & $ -1.92 + i\, 1.58 $ \\
\end{tabular}
\end{center}
\end{table}

\begin{table}
\caption{$K^- d$ scattering length (in fm) calculated with models OS1 and OS2,
in the isospin basis, and in the physical basis. SF(6.7) and 3S1 specify
the deuteron model.}
\protect\label{tab:AKmd_2}

\begin{center}
\begin{tabular}{cccc}
%
Model  &  iso-SF(6.7) & phys-SF(6.7) & phys-3S1  \\
\hline
OS1  & $ -1.76 + i\, 2.91 $ & $ -1.99 + i\, 1.65 $ & $ -1.98 + i\, 1.31 $ \\
OS2  & $ -1.37 + i\, 2.68 $ & $ -1.68 + i\, 1.66 $ & $ -1.69 + i\, 1.33 $ \\
\end{tabular}
\end{center}
\end{table}


\end{document}